\documentclass[11pt,nohyper,a4paper]{JHEP3}
\usepackage{amssymb,euscript,bbold}
\newcommand{\sect}[1]{\setcounter{equation}{0}\section{#1}}
\renewcommand{\theequation}{\arabic{section}.\arabic{equation}}
\newcommand{\bea}{\begin{eqnarray}}
\newcommand{\eea}{\end{eqnarray}}
\def\be{\begin{equation}}
\def\ee{\end{equation}}
\def\ba{\begin{eqnarray}}
\def\ea{\end{eqnarray}}
\def\nn{\nonumber \\}

\title{Supersymmetric $AdS_5$ black holes}

\author{Jan B. Gutowski\\ Mathematical Institute\\
Oxford, OX1 3LB, UK\\ E-mail:
\email{gutowski@maths.ox.ac.uk}}

\author{Harvey S. Reall\\Kavli Institute for Theoretical Physics\\
University of California, Santa Barbara   \\
CA 93106-4030, USA\\
E-mail:
\email{reall@kitp.ucsb.edu}}

\abstract{
The first examples of supersymmetric, asymptotically $AdS_5$, black
hole solutions are presented. They form a 1-parameter family of
solutions of minimal five-dimensional gauged supergravity. Their angular momentum
can never vanish. The solutions are obtained by a systematic analysis
of supersymmetric solutions with Killing horizons. Other new examples
of such solutions are obtained. These include solutions for which the
horizon is a homogeneous $Nil$ or $SL(2,R)$ manifold.}


\preprint{NSF-KITP-04-02}

\begin{document}

\setcounter{equation}{0}

\sect{Introduction}

The AdS/CFT correspondence \cite{maldacena:98}
provides a non-perturbative definition of
quantum gravity in spacetimes that are asymptotic to a product of
anti-de Sitter space with some compact manifold. In
principle, it provides a precise framework in which the puzzles of
quantum black hole physics should be solvable. The correspondence has
certainly led to an improved qualitative understanding of
black holes but quantitative results are lacking. For example, it has
not been possible to calculate precisely the entropy of, say, a
Schwarzschild-AdS black hole. The problem is that we don't know how to
compute in strongly coupled gauge theories.

The exception to these remarks is the BTZ black hole
\cite{banados:92}. In this case,
the Cardy formula of two dimensional CFT can be employed to calculate
the black hole entropy \cite{strominger:98}.
The older string theory calculations of the
entropy of asymptotically flat (nearly) supersymmetric black holes
\cite{strominger:96, breckenridge:97, maldacena:96, johnson:96, breckenridge:96,
horowitz:96, horowitz:96a} are now understood as arising from this
result, since such black holes always arise from  black string
solutions whose near-horizon geometries are locally products involving an
$AdS_3$ factor \cite{strominger:98, balasubramanian:98,
cvetic:98, cvetic:99}. The black hole entropy calculations are
applications of $AdS_3/CFT_2$ to these BTZ-like near-horizon geometries.
Hence there is a sense in which the only black holes whose
entropy has been calculated are BTZ-like solutions.

It is clearly desirable to calculate the entropy of
higher dimensional AdS black holes using the AdS/CFT correspondence.
However, in going to higher dimensions we have to
confront the problem of strong coupling in the CFT. One approach that
proved succesful in the old entropy calculations was to restrict
attention to {\it supersymmetric} black holes. Non-renormalization theorems
then allow certain results to be extrapolated from weak to strong
coupling. It is natural to ask whether this can also be done for
asymptotically AdS black holes in dimension $D>3$.

The first obstacle one encounters in attempting to do this is that it
is hard to think of any supersymmetric, asymptotically\footnote{
By ``asymptotically AdS" we mean a spacetime that is asymptotic to
global AdS rather than one which is only
asymptotically locally AdS.} AdS, black hole
solutions with $D>3$. For example, the BPS limit of the
Reissner-Nordstrom-AdS solution is a naked singularity, not a black
hole \cite{romans:92, london:95}.

A clue to overcoming this obstacle comes from the BTZ solution. The
extremal BTZ solution preserves some supersymmetry
\cite{coussaert:94}. It has mass $M = |J|
\ell$ where $J$ is the angular momentum and $\ell$ the asymptotic AdS
radius.  However, this solution only describes a black hole when $J
\ne 0$, i.e., {\it supersymmetric $AdS_3$ black holes must rotate}.
When $J=0$ it does not have a regular horizon \cite{banados:93}.

The same is true of supersymmetric, asymptotically
AdS black holes in four dimensions. If the black hole uniqueness
theorems extend to asymptotically AdS spacetimes then the most general
stationary black hole solution of $D=4$ Einstein-Maxwell theory with a negative
cosmological constant should belong to the family of charged rotating
black hole solutions obtained by Carter \cite{carter:68} (often called
Kerr-Newman-AdS solutions). The BPS limit of these solutions was examined in
\cite{kostalecky:96} in the context of minimal ${\cal N}=2$ gauged
supergravity. It was shown that supersymmetric black holes do indeed
exist. Their mass $M$ and angular momentum $J$ are uniquely determined by
their charge(s).\footnote{
There is some controversy over which charges these black holes can
carry. In \cite{kostalecky:96}, a BPS bound for this theory was
presented and it was shown that black holes saturate this bound when $M$ and
$J$ take values determined by $Q$ and $P$, the electric and magnetic
charges. This implies that there should be a 2-parameter family of
dyonic supersymmetric black holes. However, in \cite{caldarelli:99} it was
claimed that existence of a super-covariantly constant spinor implies
$P=0$, i.e., there should only be a 1-parameter family of electrically
charged supersymmetric black holes. A similar result for non-rotating
(nakedly singular) solutions was obtained in \cite{romans:92}.
These results contradict
each other since existence of a super-covariantly constant spinor
should be equivalent to saturation of the BPS bound.}
Once again, there are no supersymmetric black hole solutions with
$J=0$ (BPS black holes with small $J$ also have small $M$, and reduce to
$AdS_4$ when $J=0$). Hence all supersymmetric black holes in this
theory have non-vanishing angular momentum. This contrasts with the
ungauged theory, in which all supersymmetric black holes have
vanishing angular momentum.

These supersymmetric black holes solutions can be oxidized to $D=11$
using results of \cite{chamblin:99}. This gives supersymmetric,
asymptotically $AdS_4 \times S^7$, black hole solutions. One
might hope to be able to calculate their entropy by counting BPS operators in
the dual CFT \cite{caldarelli:00}. Unfortunately,
the dual CFT is only poorly understood
and, in particular, the spectrum of BPS operators is not known. We
shall therefore turn our attention to five dimensions.

The $AdS_5/CFT_4$ correspondence is better understood than $AdS_4/CFT_3$
because the CFT has a fairly simple
description as ${\cal N}=4$ $SU(N)$ super Yang-Mills theory. A lot is
known about BPS operators in this theory. It therefore seems a
promising arena in which to study supersymmetric black
holes. Unfortunately, there are no known supersymmetric,
asymptotically $AdS_5$, black hole solutions.\footnote{
See \cite{behrdnt:98, klemm:01, klemm:01a} for some attempts at
constructing such solutions. These attempts produced solutions with naked
singularities or naked closed timelike curves rather than black hole
solutions.}

The goal of this paper is to obtain the first examples of
supersymmetric, asymptotically $AdS_5$ black holes. Guided by the
above discussion of $D=3,4$ solutions, we should expect such
black holes to have non-zero angular momentum. That is why finding
solutions is non-trivial. Fortunately, recent advances in our
understanding of supersymmetric supergravity solutions give a
systematic method for finding such solutions.

Long ago, Tod obtained all supersymmetric solutions of the minimal
${\cal N}=2$ $D=4$ ungauged supergravity theory \cite{tod:83}.
There has recently been a revival
of interest in Tod's work, and his method has now been applied to the minimal
supergravity theories in $D=5$ \cite{gauntlett:02} and $D=6$
\cite{gutowski:03} and the
minimal gauged supergravity theories in $D=4$ \cite{caldarelli:03}
and $D=5$ \cite{gauntlett:03}
(note that all of these theories have $8$ supercharges).
The corresponding analysis has also been performed in $D=11$
\cite{gauntlett:03a, gauntlett:03b}, although
the results become increasingly less explicit as $D$ increases. The
results of \cite{gauntlett:02} were used in \cite{reall:03} to prove a
uniqueness theorem for supersymmetric black hole solutions of minimal
$D=5$ ungauged supergravity. The proof is constructive,
i.e., it yields the black hole solution explicitly. If this proof can
be adapted to other theories then it yields a systematic method for
obtaining all supersymmetric black hole solutions of such theories.

This strategy was applied to the minimal $D=6$ ungauged supergravity
in \cite{gutowski:03}. Supersymmetric solutions of this theory are, in
general, much more complicated than those of the $D=5$ theory. This
made obtaining a full uniqueness theorem too difficult. Nevertheless,
the method of \cite{reall:03} does allow one to determine
all possibilities for the near-horizon geometry of a supersymmetric
solution with a (spatially) compact event horizon.

In this paper we shall consider the minimal $D=5$ gauged supergravity
theory analyzed in \cite{gauntlett:03}. Applying the method of
\cite{reall:03} to this theory reduces finding the near-horizon
geometry to solving certain equations on a $3$-manifold $H$ corresponding
to a spacelike cross-section of the event horizon. We find several
families of near-horizon solutions in which $H$ is locally isometric to a
homogeneous metric on a group manifold, specifically\footnote{
Solutions with $H$ isometric to a homogeneous $Nil$-manifold have
been previously obtained in \cite{cadeau:01}. Unlike our solutions,
these solutions are static and not supersymmetric.}
$Nil$, $SL(2,R)$
or $SU(2)$. In the latter case, the metric is the squashed metric on
$S^3$. There is also the possibility of the near-horizon geometry
being $AdS_5$, in which case $H$ is locally isometric to $H^3$ with
its standard Einstein metric.

The most promising candidate for the near-horizon geometry of a black
hole is the solution with $H=S^3$. We therefore use the formalism of
\cite{gauntlett:03} to look for supersymmetric solutions that have
this near-horizon geometry and are asymptotically $AdS_5$. We do this
by writing both the near-horizon solution and the $AdS_5$ solution in
a canonical form discussed in \cite{gauntlett:03}. The similarity of
the two solutions suggests a natural Ansatz for obtaining a solution
that interpolates between the two. Plugging this Ansatz into the
equations of \cite{gauntlett:03} yields a 1-parameter family of
asymptotically $AdS_5$ black hole solutions preserving at least two 
of the eight supersymmetries.

Our solutions are parametrized by their (electric) charge, which
determines their mass and angular momenta. They carry equal angular
momentum in two orthogonal 2-planes, just like the supersymmetric
black holes of the ungauged $D=5$ theory \cite{breckenridge:97,
gauntlett:99}. This angular momentum vanishes only when the charge
vanishes, when the solution reduces to $AdS_5$. In this respect they
are similar to the $D=4$ solutions discussed above. Small charge black
holes are small, with low angular momentum and almost round horizons
whereas large charge black holes are large with high angular momentum
and very squashed horizons. The causal structure of our solutions is
very similar to that of the black holes of the ungauged theory, which
was analyzed in \cite{gibbons:99}. Behind the horizon there is a naked
singularity surrounded by a region of closed timelike
curves. Geodesics entering the black hole can emerge into a new
asymptotically AdS region.

Our solutions can be oxidized to give BPS solutions of type IIB supergravity
using the results of \cite{chamblin:99}. We hope that certain properties
of these supersymmetric, asymptotically $AdS_5 \times S^5$ black hole
solutions will be reliably calculable within the dual ${\cal N}=4$ $SU(N)$
super Yang-Mills theory even at weak coupling.

We should note that, although supersymmetric, asymptotically AdS,
black holes in $D=3,4,5$ must rotate, the same is not true in
$D=7$. Supersymmetric, asymptotically $AdS_7$, black hole solutions
were obtained in \cite{gauntlett:01}. These solutions are
static. However, they appear to preserve only one supersymmetry,
which is probably too few to be useful in attempting to calculate
their entropy, especially since the $AdS_7/CFT_6$ duality is not very
well understood.

This paper is arranged as follows. In section \ref{sec:susyreview} we
review the general form of supersymmetric solutions of minimal $D=5$
gauged supergravity, as deduced in \cite{gauntlett:03}. Section
\ref{sec:horizon} contains the analysis of possible near-horizon
geometries of supersymmetric solutions with horizons. In section
\ref{sec:blackhole} we obtain the supersymmetric black hole
solutions. Our results are discussed in section \ref{sec:discussion}.

The reader interested only in the black hole solutions (and not their
derivation) should jump to subsection \ref{sec:newcoord}. Other new
supersymmetric solutions are given by equations \ref{eqn:sol1},
\ref{eqn:sol2}, \ref{eqn:sol3}, \ref{eqn:limit} and \ref{eqn:limitasymp}.

\sect{Supersymmetric solutions}

\label{sec:susyreview}

The theory we shall be considering is minimal $D=5$ gauged
supergravity, with bosonic action
\be
 S = \frac{1}{4\pi G}\int \left[\left( \frac{R_5}{4} + \frac{3}{\ell^2}
 \right) \star 1 - \frac{1}{2} F \wedge \star F - \frac{2}{3\sqrt{3}} F
 \wedge F \wedge A \right],
\ee
where $R_5$ is the Ricci scalar and
$F=dA$ is the field strength of the $U(1)$ gauge field.
The bosonic equations of motion are
\bea
\label{eqofmot}
{}^5 R_{\alpha\beta}-2{}F_{\alpha\gamma}{}F_\beta{}^\gamma
+\frac{1}{3}g_{\alpha\beta}{}(F^2+{12 \over \ell^2})&=&0\nn d*{}F +
\frac{2}{\sqrt{3}} {}F \wedge {}F&=&0
\eea
where ${}F^2\equiv
{}F_{\alpha\beta}{}F^{\alpha\beta}$.
The general form of purely bosonic supersymmetric solutions of
this theory was obtained in \cite{gauntlett:03}. It was shown that
any such solution must admit certain globally defined
tensors, namely a real scalar $f$, a real vector $V$ and three real
2-forms $X^i$, $i=1,2,3$. $f$, $V$ and $X^1$ are gauge-invariant but
$X^{2,3}$ are not: $X^2 + i X^3$ picks up a phase under a gauge
transformation.

These quantities satisfy certain algebraic
relations, which are the same as in the ungauged theory
\cite{gauntlett:02}:
\be
 V^\alpha V_\alpha = - f^2,
\ee
\be
 X^i \wedge X^j =  - 2 \delta_{ij} f \star V,
\ee
\be
\label{eqn:VX}
 i_V X^i = 0,
\ee
\be
\label{eqn:Xasd}
 i_V \star X^i = - f X^i,
\ee
\be
\label{eqn:Xalg}
 \left( X^i \right)_{\gamma \alpha} \left( X^j \right)^\gamma {}_\beta
= \delta_{ij} \left( f^2 \eta_{\alpha \beta} + V_\alpha V_\beta
\right) - \epsilon_{ijk} f \left( X^k \right)_{\alpha \beta},
\ee
where $\epsilon_{123}=+1$ and, for a $p$-form $\Omega$ and vector $Y$,
$i_Y \Omega$ denotes the $(p-1)$-form obtained by contracting $Y$ with
the first index of $\Omega$. From the first expression we see that $V$
is timelike or null (the possibility of $V$ vanishing anywhere
can be excluded using an argument in \cite{reall:03}).

There are also differential relations \cite{gauntlett:03}:
\be
\label{eqn:df}
 df = -\frac{2}{\sqrt{3}} i_V F,
\ee
\be
\label{eqn:VKilling}
 D_{(\alpha} V_{\beta)} = 0,
\ee
\be
\label{eqn:dV}
 dV = -\frac{4}{\sqrt{3}} f F - \frac{2}{\sqrt{3}} \star \left( F
\wedge V \right) - 2 \ell^{-1} X^1
\ee
and
\be
\label{eqn:dX}
 dX^i = \frac{1}{\ell} \epsilon_{1ij} \left[ 2 \sqrt{3} A \wedge X^j + 3
\star X^j \right],
\ee
hence $X^1$ is closed, but $X^{2,3}$ need not be.
Equations (\ref{eqn:VKilling}) and (\ref{eqn:df}) imply that $V$ is a
Killing vector field that leaves the Maxwell field strength invariant
\cite{gauntlett:03}, i.e., $V$ generates a symmetry of the solution.

There are two cases to consider. In the first case, $V$ is globally
null, i.e., $f$ vanishes everywhere. A general analysis of such
solutions was presented in \cite{gauntlett:03} where it was shown that
they preserve at least $1/4$ of the supersymmetry. We shall refer to
these as the null family of solutions.

We shall be primarily interested in the case in which $V$ is not
globally null. We shall refer to such solutions as belonging to the
timelike family because for these solutions we can always find some open set
${\cal U}$ in which $V$ is timelike. There is no loss of generality in
assuming $f>0$ in ${\cal U}$. Coordinates can be introduced so that
the metric in ${\cal U}$ takes the form
\be
\label{eqn:metric}
 ds^2 = -f^2 (dt + \omega)^2 + f^{-1} ds_4^2,
\ee
where $V = \partial/\partial t$, $ds_4^2$ is the line element of a
four-dimensional Riemannian ``base space" ${\cal B}$ orthogonal to the
orbits of $V$, and $\omega$ is a $1$-form on ${\cal B}$. The strategy
of \cite{gauntlett:03} (following \cite{gauntlett:02}) is to reduce
necessary and sufficient conditions for supersymmetry to a set of
equations on ${\cal B}$.

Equation (\ref{eqn:VX}) implies that the $2$-forms $X^i$ can be regarded as
$2$-forms on ${\cal B}$, and equation (\ref{eqn:Xasd}) implies that
these are anti-self dual if we take the volume form $\eta_4$ of ${\cal
B}$ to be related to the five-dimensional volume form $\eta$ by
\be
 \eta_4 = f i_V \eta.
\ee
Equation (\ref{eqn:Xalg}) implies that the $2$-forms obey the algebra of
the imaginary unit quaternions on ${\cal B}$, i.e., they define an
almost hyper-K\"ahler structure. This is not integrable: supersymmetry
merely requires that ${\cal B}$ is K\"ahler, with K\"ahler form $X^1$
\cite{gauntlett:03}.

Equations (\ref{eqn:df}) and (\ref{eqn:dV}) can be solved to determine the
field strength, giving \cite{gauntlett:03}
\be
\label{eqn:maxwell1}
 F = \frac{\sqrt{3}}{2} d\left[ f (dt + \omega) \right] -
 \frac{1}{\sqrt{3}} G^+ - \sqrt{3} \ell^{-1} f^{-1} X^1,
\ee
where $G^{\pm}$ is defined by
\be
 G^\pm = \frac{f}{2} \left( d\omega \pm \star_4 d\omega \right),
\ee
and $\star_4$ denotes the Hodge dual on ${\cal B}$. Supersymmetry
requires that \cite{gauntlett:03}
\be
\label{eqn:fsol}
 f^{-1} = -\frac{\ell^2 R}{24},
\ee
where $R$ is the Ricci scalar of ${\cal B}$, and
\be
\label{eqn:Geq}
 G^+ = -\frac{\ell}{2} \left( {\cal R} - \frac{R}{4} X^1 \right)
\ee
where ${\cal R}$ is the Ricci form on ${\cal B}$, defined by
($m,n,\ldots$ denote curved indices on ${\cal B}$)
\be
 {\cal R}_{mn} = \frac{1}{2} R_{mnpq} \left( X^1 \right)^{pq},
\ee
with $R_{mnpq}$ the Riemann tensor of ${\cal B}$.

The final condition arises from the Maxwell equation, which gives
\cite{gauntlett:03}
\be
 \label{eqn:maxwell}
 \nabla^2 f^{-1} = \frac{2}{9} (G^+)^{mn} (G^+)_{mn} + \ell^{-1} f^{-1}
(G^-)^{mn} (X^1)_{mn} - 8 \ell^{-2} f^{-2},
\ee
where $\nabla^2$ is the Laplacian on ${\cal B}$.
These conditions are necessary for supersymmetry. It turns out that
they are also sufficient, with the solution preserving at least $1/4$ of the
supersymmetry \cite{gauntlett:03}.\footnote{
It was claimed in \cite{gauntlett:03} that these solutions are $1/2$
supersymmetric but examining the projections imposed on the spinor
reveals that they are actually $1/4$ supersymmetric.}
Hence all supersymmetric solutions in
the timelike family are determined as follows. First pick a K\"ahler
manifold ${\cal B}$ with negative Ricci scalar.
Equation (\ref{eqn:fsol}) determines $f$. Now find
a $1$-form $\omega$ so that equations (\ref{eqn:Geq}) and (\ref{eqn:maxwell})
are satisfied (where $X^1$ is the K\"ahler form). It was demonstrated
in \cite{gauntlett:03} that a solution always exists.
Then there will be some region ${\cal U}$
of spacetime in which the metric is given by (\ref{eqn:metric}) and the
field strength by (\ref{eqn:maxwell1}), which can also be written
\be
 F = \frac{\sqrt{3}}{2} d \left[ f (dt + \omega) \right] +
\frac{\ell}{2\sqrt{3}} {\cal R}.
\ee
The solution outside of ${\cal U}$ can then be obtained by analytic
continuation.

Note that the only information used in deriving the general
supersymmetric solution (in either family) is the equations satisfied
by $f$, $V$ and $X^i$. Hence these equations are both necessary and
sufficient for supersymmetry.

We should mention that some solutions belong to both the timelike and
null families. This is only possible for solutions preserving more
than $1/4$ of the supersymmetry. Examples are the maximally
supersymmetric $AdS_5$ solution, and the $AdS_3 \times H^2$ solution
\cite{gauntlett:03}. Since $AdS_5$ is the unique maximally
supersymmetric solution \cite{gauntlett:03}, the latter solution
presumably preserves $1/2$ or $3/4$ of the supersymmetry.

\sect{Near horizon analysis}

\label{sec:horizon}

\subsection{General analysis}

If a supersymmetric solution has a physical horizon ${\cal H}$ then it
must be preserved by all Killing vectors of the spacetime and in
particular by the supersymmetric Killing vector field $V$. This
implies that $V$ must be spacelike or null on the horizon. However, we
know that $V$ cannot be spacelike hence $V$ must be null on the
horizon, i.e., the horizon is a Killing horizon of $V$.

If a timelike solution has a horizon then ${\cal H}$ cannot intersect
${\cal U}$ because $V$ is null on ${\cal H}$ but timelike in ${\cal
U}$. Therefore it is not possible to identify which solutions have
horizons just by looking at the solution in the form presented
above. We shall instead adopt the approach introduced in
\cite{reall:03}. This approach is equally valid for timelike or null
solutions.

The idea is to introduce a coordinate system adapted to the presence
of a Killing horizon, namely Gaussian null coordinates, and then to
examine the equations satisfied by $f$, $V$ and $X^i$ in such
coordinates (recall that these are necessary and sufficient for
supersymmetry). In the near-horizon limit, they reduce to equations on a
$3$-manifold $H$ corresponding to a constant time slice through ${\cal
H}$ (i.e. $H = {\cal H} \cap \Sigma$ where $\Sigma$ is a
spacelike slice that intersects ${\cal H}$). In the ungauged theory
considered in \cite{reall:03} it was shown that these equations can be
completely solved when $H$ is compact and the near-horizon geometry
can therefore be determined explicitly. This has also been done in the
minimal six-dimensional ungauged supergravity \cite{gutowski:03}.

In Gaussian null coordinates, the line element takes the form \cite{reall:03}
\be
\label{eqn:metric2}
 ds^2 = -r^2 \Delta^2 du^2 + 2 du dr + 2 r h_A du dx^A + \gamma_{AB}
dx^A dx^B
\ee
with
\be
 V = \frac{\partial}{\partial u}, \qquad f = r\Delta.
\ee
The horizon is at $r=0$, and $H$ is given by $r=0$ and $u={\rm
constant}$. The region exterior to the horizon is $r>0$, where we can
assume $\Delta \ge 0$. The quantities $\Delta$, $h_A$ and
$\gamma_{AB}$ depend smoothly on $r$ and $x^A$ but are independent of
$u$ (because $V$ is Killing).

The near-horizon limit is defined by
$r=\epsilon \tilde{r}$ and $u = \tilde{u}/\epsilon$ with $\epsilon
\rightarrow 0$. After taking this limit we recover a metric of the
same form but with $\Delta$, $h_A$ and $\gamma_{AB}$ depending only on
$x^A$. This is the reason why obtaining the near-horizon geometry of a
supersymmetric solution reduces to solving equations on $H$. In what
follows, we shall {\it not} assume that the near-horizon limit has
been taken since it is not immediately obvious that this limit must
preserve supersymmetry. We shall instead evaluate all equations as a
power series in $r$. The near-horizon limit corresponds to
discarding all but the ${\cal O}(r^0)$ terms.

We are free to choose a gauge in which the gauge field $A$ is a smooth
function of $r$ with
\be
 A_u \equiv i_V A = \frac{\sqrt{3}}{2} f.
\ee
In this gauge, we have ${\cal L}_V A = 0$ (using equation
(\ref{eqn:df})). Write
\be
 A = \frac{\sqrt{3}}{2} r \Delta du + A_r dr + a_A dx^A.
\ee
In general, $a_A$ will not be globally defined on $H$. Note also that
taking the near-horizon limit removes $A_r$.

The 2-forms $X^i$ can be written as \cite{reall:03}
\be
 X^i = dr \wedge Z^i + r \left( h \wedge Z^i - \Delta \star_3 Z^i \right),
\ee
where $h\equiv h_A dx^A$,
$\star_3$ denotes the Hodge dual with respect to $\gamma_{AB}$
(the volume form $\eta_3$ of $\gamma_{AB}$ is chosen so that $du \wedge dr
\wedge \eta_3$ has positive five-dimensional orientation), and
\be
 Z^i = Z^i_A dx^A
\ee
are a set of vector fields orthonormal with respect to $\gamma_{AB}$,
i.e.,
\be
\label{eqn:starZ}
 \star_3 Z^i = \frac{1}{2} \epsilon_{ijk} Z^j \wedge Z^k.
\ee
$Z^1$ is gauge-invariant and hence globally defined on $H$. However,
$Z^{2,3}$ are gauge-dependent: if we define
\be
 W \equiv Z^2 + iZ^3
\ee
then $W$ picks up a phase under a gauge transformation.
Equation (\ref{eqn:dX}) gives
\bea
 \hat{d}Z^i &=& h \wedge Z^i - \Delta \star_3 Z^i +r \partial_r \big(h
\wedge Z^i - \Delta \star_3 Z^i \big)
\nn
 &+& \ell^{-1} \epsilon_{1ij} \left[ 3 \star_3 Z^j +
2\sqrt{3} a \wedge Z^j - 2\sqrt{3} r A_r \left( h \wedge Z^j - \Delta
\star_3 Z^j \right) \right],
\eea
 and also
\bea
\label{eqn:dhfull}
 \star_3 \,\hat{d} h &=& {\hat{d}} \Delta + \Delta h -r (\partial_r \Delta) h
+2r\Delta \partial_r h +r \star_3 (h \wedge \partial_r h)+r \Delta^2 \epsilon_{ijk} Z^i
\langle Z^j , \partial_r Z^k \rangle
\nn
&+& \frac{\Delta}{\ell} \left( 6 + 4 \sqrt{3} r
A_r \Delta \right) Z^1.
\eea
In these equations, $\langle,\rangle$ denotes the inner product defined by
$\gamma_{AB}$, and $\hat{d}$ is defined by
\be
 (\hat{d} Y)_{ABC\ldots} = (p+1)\partial_{[A} Y_{BC\ldots]}
\ee
for any $p$-form $Y$ with only $A,B,C,\ldots$ indices.
To leading order in $r$, these equations reduce to
\be
\label{eqn:dZ}
 \hat{d} Z^i = - \Delta \star_3 Z^i + h
\wedge Z^i + \ell^{-1} \epsilon_{1ij} \left( 3 \star_3 Z^j + 2\sqrt{3} a
\wedge Z^j \right) + {\cal O}(r)
\ee
and
\be
\label{eqn:dh}
 \star_3 \, \hat{d} h - \hat{d} \Delta - \Delta h = \frac{6 \Delta}{\ell}
 Z^1 + {\cal O}(r).
\ee
We shall be interested in evaluating such equations at $r=0$, i.e., on
the $3$-manifold $H$. From equation (\ref{eqn:starZ}) we obtain
\be
\label{eqn:dstarZ}
 \hat{d}^\dagger Z^i = 2 \left( h \cdot Z^i + 3\ell^{-1}
 \delta_{i1} \right) + 2\sqrt{3} \ell^{-1} \epsilon_{1ij} a \cdot Z^j +
{\cal O}(r),
\ee
where $\hat{d}^\dagger \equiv \star_3 \hat{d} \star_3$, and
$h \cdot Z^i$ and $a \cdot Z^j$ are defined by contracting
indices with $\gamma^{AB}$.

For $r>0$, if $\Delta > 0$ then $V$ is timelike and hence $G^+$ is
well-defined. Self-duality implies that $G^+$ can be written \cite{reall:03}
\be
 G^+ = dr \wedge {\cal G} + r \left( h \wedge {\cal G} + \Delta
\star_3 {\cal G} \right),
\ee
where ${\cal G} \equiv {\cal G}_A dx^A$. We obtain
\bea
 {\cal G} &=& -{3 \over 2 r \Delta^2} {\hat{d}} \Delta +{3 \over 2
\Delta^2} (\partial_r \Delta) h -{3 \over 2 \Delta} \partial_r h -{1
\over 2} \epsilon_{ijk} Z^i \langle Z^j , \partial_r Z^k \rangle
\nn
&-& \frac{1}{2 r \ell \Delta} \left( 6 + 4 \sqrt{3} r
A_r \Delta \right) Z^1,
\eea
where we have used equation (\ref{eqn:dhfull}) to eliminate $\hat{d}
h$. We can now determine the Maxwell field strength from equation (\ref{eqn:maxwell1}):
\bea
 F &=& {\sqrt{3} \over 2} \big[
 - \partial_r(r \Delta) du \wedge dr -r du \wedge {\hat{d}} \Delta
 +{1 \over 3} \epsilon_{ijk} dr \wedge Z^i  \langle Z^j , \partial_r Z^k \rangle
 \nn
 &-& \star_3 h -r \star_3 \partial_r h +{r \over 3}(-2 \Delta \star_3 Z^i +h \wedge Z^i)
 \langle Z^j , \partial_r Z^k \rangle \big]
\nn
&+& \ell^{-1} \left[2 A_r \left( dr \wedge Z^1 + r h \wedge Z^1
 \right) - \left( \sqrt{3} + 4 r A_r \Delta \right) \star_3 Z^1
 \right].
\eea
Once again, we have used equation (\ref{eqn:dhfull}) to eliminate $\hat{d} h$.
Note that $F$ is regular at $r=0$ even if $\Delta = 0$.
Despite the appearance of the gauge-dependent quantity $A_r$, this expression
is gauge invariant because there are terms involving
$Z^{2,3}$, which are also gauge-dependent. Under a gauge
transformation, the gauge dependence of these different quantities must
cancel. In any case, we shall only be interested in evaluating $F$ at
$r=0$ where gauge-independence is manifest. For example, the $AB$
component of $F$ is given by
\be
 \frac{1}{2} F_{AB}dx^A \wedge dx^B = - \frac{\sqrt{3}}{2} \star_3
 \left( h + 2\ell^{-1} Z^1 \right) + {\cal O}(r),
\ee
hence
\be
\label{eqn:da}
 \hat{d} a = - \frac{\sqrt{3}}{2} \star_3
 \left( h + 2\ell^{-1} Z^1 \right) + {\cal O}(r).
\ee
The $ABC$ component of the Bianchi identity for $F$ is therefore
\be
 \hat{d} \left[ \star_3 \left( h + 2 \ell^{-1}  Z^1 \right) \right] =
{\cal O}(r).
\ee
Combining this with equation (\ref{eqn:dstarZ}) gives
\be
\label{eqn:dstarh}
 \hat{d}^\dagger h = -4 \ell^{-1} \left( h \cdot Z^1 + 3 \ell^{-1}
\right) + {\cal O}(r).
\ee
Note that this implies that $h$ cannot be identically zero on $H$.
The next step of \cite{reall:03} is to consider
\be
\label{eqn:ddelta}
 \hat{d} \Delta \wedge \star_3 \, \hat{d} \Delta = \hat{d} \left(
\Delta \hat{d} h - \frac{\Delta^2}{2} \star_3 h - 3 \ell^{-1} \Delta^2
\star_3 Z^1 \right) + 4 \ell^{-1} \Delta^2 \left( h \cdot Z^1 + 3
\ell^{-1} \right) + {\cal O}(r).
\ee
where we have used equations (\ref{eqn:dh}), (\ref{eqn:dstarZ}) and
(\ref{eqn:dstarh}). The expression inside the bracket is globally
defined on $H$. Hence in the ungauged theory ($\ell=\infty$) the right
hand side is exact, so integrating this expression over $H$ implies
that $\Delta$ is constant on $H$ if $H$ is compact. However this
simple argument is no longer possible in the gauged theory owing to
the presence of the final term above.

We can calculate the spin connection using equation (\ref{eqn:dZ}). This
gives
\ba
\label{eqn:gradZ}
 \nabla_A Z^i_B = -\frac{\Delta}{2} \left( \star_3 Z^i \right)_{AB} +
 \gamma_{AB} \left( h \cdot Z^i + 3 \ell^{-1}
 \delta_{i1} \right) - Z^i_A h_B - 3 \ell^{-1} Z^i_A Z^1_B  \nn
 + 2 \sqrt{3} \ell^{-1} \epsilon_{1ij} a_A Z^j_B + {\cal O}(r),
\ea
where $\nabla$ is the connection associated with $\gamma_{AB}$. The
Ricci tensor $R_{AB}$ of $\gamma_{AB}$ can now be obtained using
\be
 R_{AB} Z^{iB} = \nabla_B \nabla_A Z^{iB} - \nabla_A \nabla_B Z^{iB}.
\ee
All gauge dependent terms cancel, as they must, giving
\be
 R_{AB} = \left( \frac{\Delta^2}{2} + h
 \cdot h + 4 \ell^{-1} h \cdot Z^1
 \right) \gamma_{AB} - h_A h_B - \nabla_{(A} h_{B)} - 6 \ell^{-1} h_{(A}
 Z^1_{B)} - 6 \ell^{-2} Z^1_A Z^1_B + {\cal O}(r).
\ee
So far we have not assumed that $H$ is compact. In the ungauged
theory, it is possible to exploit compactness of $H$ to prove that
$h$ must be a Killing vector field on $H$ \cite{reall:03}.
In the gauged theory, it is natural to guess that if such a Killing
vector field exists then it will be a linear combination of $h$ and
$Z^1$. To look for such a Killing vector field, it is natural to
define
\be
 N = h + 2 \ell^{-1} Z^1,
\ee
which is coclosed on $H$. Following the strategy of
\cite{reall:03}, we look for a value of $\alpha$ such that $N + \alpha
Z^1$ is Killing. To do this, define
\be
 I \equiv \int_H \nabla_{(A} (N+\alpha Z^1)_{B)} \nabla^{(A} (N+\alpha
Z^1) ^{B)}.
\ee
We need to find a value for $\alpha$ which gives $I=0$. $I$ can be
calculated by integration by parts, and commuting derivatives using
the above expression for $R_{AB}$. We find
\be
\label{eqn:Iidenta}
 \int_H \nabla_{(A} N_{B)} \nabla^{(A} N^{B)} = \int_H \left[ \frac{2}{\ell}
 N \cdot Z^1 N \cdot N + \frac{4}{\ell^2} N \cdot N - \frac{2}{\ell^2} (N
\cdot Z^1)^2 \right],
\ee
\bea
\label{eqn:altident}
 \int_H \nabla_{(A} Z^1_{B)} \nabla^{(A} N^{B)} &=& \int_H \big[\frac{2}{\ell} N
\cdot N -
\frac{1}{2\ell} (N \cdot Z^1)^2 - \frac{2}{ \ell^3}
\nn &-& {\frac{1}{2}} N\cdot Z^1
N \cdot N  -
\Delta^2 \left( N \cdot Z^1 + \frac{1}{\ell} \right) \big]
\nn
&=& \int_H \left[ {1 \over 2} (N \cdot N) (N \cdot Z^1) +{5 \over 2
\ell} (N \cdot Z^1)^2-{2 \over \ell^3} \right]
\eea
\be
\label{eqn:Iidentc}
 \int_H \nabla_{(A} Z^1_{B)} \nabla^{(A} Z^{1B)} = \int_H
\left[\frac{1}{2} N \cdot N + \frac{3}{2} ( N \cdot Z^1)^2 - \frac{2}{ \ell^2}
\right].
\ee
The first expression in (\ref{eqn:altident}) is obtained by using
integration by parts to remove the derivative on $N$. The second
expression is obtained by substituting in equation (\ref{eqn:gradZ}) and
then integrating some terms by parts.

Note that ({\ref{eqn:altident}}) together with ({\ref{eqn:ddelta}}) implies that
\be
\label{eqn:identthree}
\int_H  \hat{d} \Delta \wedge \star_3 \, \hat{d} \Delta
= 4 \ell^{-1} \int_H \Delta^2 ((N \cdot Z^1)+ \ell^{-1}) = 4 \ell^{-1}
\int_H {2 \over \ell} N \cdot N -{3 \over \ell} (N \cdot Z^1)^2
- N \cdot Z^1 N \cdot N.
\ee
Unfortunately, using only the above expressions for the integrals,
there is no value for $\alpha$ that guarantees $I=0$. In the ungauged
theories considered in \cite{reall:03} and \cite{gutowski:03} it was
shown that, when $H$ is compact, $\Delta$ must be constant on $H$ and
that $H$ must admit a Killing vector. We have been unable to establish
the same results in the gauged theory. This prevents us from classifying
all possible near-horizon solutions. Nevertheless, we shall now see
that it is still posssible to use the above results to obtain
particular examples of near-horizon solutions.

\subsection{Near-horizon solutions}

In this subsection, we shall be interested in the near-horizon
solution. As mentioned above, this amounts to neglecting the ${\cal
O}(r)$ terms, which we shall do henceforth.
Determining the near-horizon geometry is equivalent to finding a
solution $(Z^i,\Delta,h,a)$ of equations (\ref{eqn:starZ}),
(\ref{eqn:dZ}), (\ref{eqn:dh}) and (\ref{eqn:da}) on $H$. The other
equations are consequences of these ones.
Unfortunately, we have not been able to find the general solution of
these equations even for compact $H$.
We shall resort to additional assumptions to obtain solutions.

Guided by the results of \cite{reall:03}, we
shall first assume that $\Delta$ is constant and non-zero on $H$.
Equation (\ref{eqn:dh}) then implies
\be
 \hat{d}^\dagger \left(h + 6 \ell^{-1} Z^1 \right) = 0.
\ee
Combining this with equations (\ref{eqn:dstarZ}) and (\ref{eqn:dstarh})
implies
\be
 h \cdot Z^1 = -\frac{3}{\ell}, \qquad N \cdot Z^1 = -\frac{1}{\ell}.
\ee
Assuming $H$ is compact, substituting $N \cdot Z^1 = -\ell^{-1}$
into ({\ref{eqn:identthree}}) we find that
\be
\int_H N \cdot N - (N \cdot Z^1)^2 =0
\ee
and hence
\be
\label{eqn:hZ}
 h = - \frac{3}{\ell} Z^1.
\ee
Hence from ({\ref{eqn:gradZ}}) it follows that $Z^1$ is Killing, and hence $h$ is Killing.

Conversely, instead of assuming that $\Delta$ is constant, we could
assume that $h$ is Killing on $H$, and $H$ is compact.
Then $\hat{d}^\dagger h = 0$, so from (\ref{eqn:dstarh}) we find
$N \cdot Z^1 = -\ell^{-1}$ and hence from  ({\ref{eqn:identthree}})
we can conclude that $\Delta$ is
constant. Equation
(\ref{eqn:hZ}) then follows as above if $\Delta \ne 0$, and it is also easily
derived if $\Delta=0$.

Another starting point would be to assume that $h = \alpha Z^1$ for
some constant $\alpha$. Equations (\ref{eqn:dstarh}) and
(\ref{eqn:dstarZ}) then imply $\alpha = -3/\ell$ or $\alpha=-2/\ell$. In the
former case, it follows from equation \ref{eqn:gradZ} that $Z^1$ and $h$
are Killing, which is the case just discussed. At
the end of this subsection, we shall show that the latter case gives
an $AdS_5$ near-horizon geometry.

Motivated by these results, we shall now construct the near-horizon
geometry of any supersymmetric solution that has $\Delta$ constant on
$H$, and $h,Z^1$ Killing on $H$ with $h=-(3/\ell)Z^1$. We shall not
assume that $H$ is compact or that $\Delta$ is non-zero on $H$.

Equation (\ref{eqn:dZ}) implies (without any assumptions)
\be
\label{eqn:dW}
 dW = \left[ (-i \Delta + 3 \ell^{-1} ) Z^1 + h - 2 \sqrt{3} i \ell^{-1} a
\right] \wedge W,
\ee
hence $W \wedge dW = 0$ on $H$ so locally we can write
\be
 W = \sqrt{2} \lambda dw,
\ee
where $\lambda$ and $w$ are locally defined complex functions on $H$. $w$ is
gauge-invariant but $\lambda$ is not. Locally, we can always perform a
gauge transformation to make $\lambda$ real, which we shall assume
henceforth.

We can introduce real functions $x,y$ by writing $w =
2^{-1/2}(x+iy)$. Note that $Z^1$ is orthogonal to $dx$ and $dy$ so the
integral curves of $Z^1$ lie within surfaces of constant $x$ and
$y$. Hence we can define coordinates $(x,y,z)$ where $z$ is the
parameter along the integral curves of $Z^1$, i.e., $Z^1 =
\partial/\partial z$. The metric on $H$ must take the form
\be
 ds_3^2 = (dz + \alpha)^2 + 2 \lambda^2 dw d\bar{w},
\ee
where $\alpha = \alpha_w dw + \alpha_{\bar{w}} d\bar{w}$. The
assumption that $Z^1$ is Killing implies that $\alpha$ and $\lambda$
are independent of $z$. We also have
\be
 h = -3\ell^{-1} (dz + \alpha).
\ee
Equation (\ref{eqn:dZ}) with $h \propto Z^1$ becomes
\be
 \partial_w \alpha_{\bar{w}} - \partial_{\bar{w}} \alpha_w = -i \Delta
\lambda^2.
\ee
This equation determines $\alpha$ up to a gradient which can be
absorbed into the definition of $z$. Equation (\ref{eqn:dW}) can be
solved for $a$, giving
\be
 a = - \frac{\Delta \ell}{2\sqrt{3}} ( dz + \alpha) -
\frac{i\ell}{2\sqrt{3}} \left( \frac{ \partial_w \lambda}{\lambda} dw -
\frac{\partial_{\bar{w}} \lambda}{\lambda} d\bar{w}\right).
\ee
Substituting this into equation (\ref{eqn:da}) gives
\be
 \partial_w \partial_{\bar{w}} \log \lambda = \frac{1}{2}
\left(3\ell^{-2} - \Delta^2 \right) \lambda^2.
\ee
This is Liouville's equation. There are three cases to consider.

The first case is $\Delta = \sqrt{3} /\ell$. We then have $\log \lambda = {\cal{F}}(w) +
\bar{{\cal{F}}}(\bar{w})$ where ${\cal{F}}$ is an arbitrary holomorphic function. A
holomorphic change of coordinate $w \rightarrow w'(w)$ can be used to
set ${\cal{F}} = 0$, i.e., $\lambda = 1$. Solving for $\alpha$ gives
\be
 \alpha = \frac{\sqrt{3}}{2 \ell} (y dx - x dy),
\ee
hence the metric of $H$ is
\be
 ds_3^2 = \left(dz + \frac{\sqrt{3}}{2 \ell} (y dx - x dy) \right)^2 +
dx^2 + dy^2,
\ee
the standard homogeneous metric on the $Nil$ group manifold. The
near-horizon limit of such a solution is
\ba
\label{eqn:sol1}
 ds^2 &=& - \frac{3 r^2}{\ell^2} du^2 + 2 du dr - \frac{6r}{\ell}du \left(dz
+ \frac{\sqrt{3}}{2 \ell} (y dx - x dy) \right) \nn &+& \left(dz +
\frac{\sqrt{3}}{2 \ell} (y dx - x dy) \right)^2 + dx^2 + dy^2 \nn
F &=& -\frac{3}{2\ell}du \wedge dr + \frac{\sqrt{3}}{2\ell} dx \wedge dy.
\ea
Note that dimensional reduction on $\partial/\partial z$ gives an
$AdS_2 \times R^2$ geometry.

The second case is $0 \le \Delta < \sqrt{3} /\ell$. The general solution to the
Liouville equation is
\be
 \lambda^2 = \frac{2 {\cal{F}}'(w) \bar{{\cal{F}}}'(\bar{w})}{({\cal{F}}(w) +
\bar{{\cal{F}}}(\bar{w}))^2 (3 \ell^{-2} - \Delta^2)},
\ee
where ${\cal{F}}$ is an arbitrary holomorphic function. A holomorphic change
of coordinates $w \rightarrow w'(w)$ can be used to set ${\cal{F}}=w$, i.e.,
\be
 \lambda^2 = \frac{1}{(3 \ell^{-2} - \Delta^2) x^2}.
\ee
Solving for $\alpha$ gives
\be
 \alpha = \frac{\Delta}{(3 \ell^{-2} - \Delta^2)} \frac{dy}{x},
\ee
hence the metric of $H$ is
\be
 ds_3^2 = \left(dz + \frac{\Delta}{(3 \ell^{-2} - \Delta^2)} \frac{dy}{x}
\right)^2 + \frac{1}{3 \ell^{-2} - \Delta^2} \left( \frac{dx^2+dy^2}{x^2}
\right),
\ee
the standard homogeneous Riemannian metric on the $SL(2,R)$ group
manifold, except for $\Delta=0$ when it degenerates to $R \times
H^2$. Taking the near-horizon limit gives the solution
\ba
\label{eqn:sol2}
 ds^2 &=& -r^2 \Delta^2 du^2 + 2 du dr - \frac{6r}{\ell} du \left(dz +
\frac{\Delta}{(3 \ell^{-2} - \Delta^2)} \frac{dy}{x} \right) \nn
 &+& \left(dz + \frac{\Delta}{(3 \ell^{-2} - \Delta^2)} \frac{dy}{x}
\right)^2 + \frac{1}{3 \ell^{-2} - \Delta^2} \left( \frac{dx^2+dy^2}{x^2}
\right) \nn
 F &=& -\frac{\sqrt{3}}{2} \Delta du \wedge dr + \frac{\sqrt{3}}{2\ell (3
\ell^{-2} - \Delta^2) x^2} dx \wedge dy,
\ea
where $\Delta$ is constant everywhere. Dimensional reduction on
$\partial/\partial z$ yields an $AdS_2 \times H^2$ geometry. If
$\Delta=0$ then the five dimensional geometry is $AdS_3 \times
H^2$, the near-horizon geometry of the black strings of
\cite{klemm:00}.

The third case is $\Delta > \sqrt{3}/\ell$. Solving the Liouville equation gives
\be
 \lambda^2 = \frac{2 {\cal{F}}'(w) \bar{{\cal{F}}}'(\bar{w})}{(1+{\cal{F}}(w)
\bar{{\cal{F}}}(\bar{w}))^2(\Delta^2 - 3 \ell^{-2}) },
\ee
where ${\cal{F}}$ is an arbitrary holomorphic function. A holomorphic change
of coordinate can be used to set ${\cal{F}}=w$. It is then convenient to
introduce real coordinates $\theta,\phi$ defined by
\be
 w = \tan (\theta/2) e^{i\psi},
\ee
giving
\be
 \lambda^2 = \frac{2 \cos^4 (\theta/2) }{\Delta^2 - 3 \ell^{-2}},
\ee
\be
 \alpha = \frac{\Delta}{\Delta^2 - 3 \ell^{-2}} \cos \theta d \psi,
\ee
and the metric on $H$ is
\be
 ds_3^2 = \frac{1}{\Delta^2 - 3 \ell^{-2}} \left[
\frac{\Delta^2}{\Delta^2 - 3 \ell^{-2}} \left( d\phi'' + \cos \theta d \psi
\right)^2 + d\theta^2 + \sin^2 \theta d\psi^2 \right],
\ee
where $\phi''$ is defined by
\be
 z = \frac{\Delta}{\Delta^2 - 3\ell^{-2}} \phi''.
\ee
We have included the primes to avoid confusion with other coordinates
$\phi,\phi'$ to be introduced later.
We see that $H$ has the homogeneous geometry of a squashed $S^3$,
i.e. the $SU(2)$ group manifold. Large $\Delta$ corresponds to a small
almost round $S^3$ whereas $\Delta$ close to $\sqrt{3}/\ell$ corresponds
to a large, highly squashed $S^3$. The near horizon solution is
\ba
\label{eqn:sol3}
 ds^2 &=& - r^2 \Delta^2 du^2 + 2 du dr - \frac{6\Delta r}{\ell (\Delta^2
- 3 \ell^{-2})} du (d \phi'' + \cos \theta d\psi ) \nn
   &+& \frac{1}{\Delta^2 - 3 \ell^{-2}} \left[ \frac{\Delta^2}{\Delta^2 -
3 \ell^{-2}} \left( d\phi'' + \cos \theta d \psi \right)^2 + d\theta^2 +
\sin^2 \theta d\psi^2 \right] \nn
 F &=& - \frac{\sqrt{3}}{2} \Delta du \wedge dr + \frac{\sqrt{3} \sin
\theta}{ 2 \ell (  \Delta^2 - 3 \ell^{-2} ) } d \theta \wedge d\psi,
\ea
where $\Delta$ is constant everywhere. Dimensional reduction on
$\partial/\partial \phi''$ yields an $AdS_2 \times S^2$ geometry.

In each of these cases, the metric on $H$ is that of a homogeneous
metric on a group manifold. This is a purely local result - the global
topology of $H$ may differ from that of the group manifold by discrete
identifications if this can be done consistently with
supersymmetry.

Finally, we mentioned above the possibility of solutions with $h =
-(2/\ell) Z^1$ on $H$. In this case, it is easy to see that equations
(\ref{eqn:dZ}) and (\ref{eqn:dh}) imply $\Delta = 0$ and $dZ^1=0$ on
$H$. Hence there is some locally defined function $z$ on $H$ such that
$Z^1 = dz$. Equation (\ref{eqn:da}) shows that we can choose the gauge
$a=0$ on $H$, then (\ref{eqn:dW}) implies $\exp (-z/\ell) W$ is closed on
$H$ so locally there is some complex function $w$ on $H$ such that $W
= \sqrt{2} \exp (z/\ell) dw$. It follows that the metric on $H$ is
\be
 ds_3^2 = dz^2 + 2 e^{2z/\ell} dw d\bar{w},
\ee
the standard Einstein metric on $H^3$. Note that this is an example of
a solution for which $h$ and $Z^1$ are not Killing. Taking the
near-horizon limit of such a solution gives the solution
\ba
\label{eqn:sol4}
 ds^2 &=& 2 du dr - \frac{4r}{\ell} du dz + dz^2 +  2 e^{2z/\ell} dw
d\bar{w} \nn
 F &=& 0,
\ea
which is just $AdS_5$ with vanishing electromagnetic
field. Written in this form, the solution belongs
to the null family.

\subsection{The base space}

If we view the above near-horizon geometries as solutions in their own
right then, for $r>0$, $V$ is timelike and the solutions belong to the
timelike class of \cite{gauntlett:03} (with the exception of the
solutions with $\Delta=0$, which are in the null class) and are
therefore supersymmetric. In general, the base space of the metric
(\ref{eqn:metric2}) has metric
\be
 ds_4^2 = r \Delta \gamma_{AB} dx^A dx^B + \frac{r}{\Delta}\left(
\frac{dr}{r} + h_A dx^A \right)^2.
\ee
For the above solutions, this is
\be
 ds_4^2 = d\rho^2 + \frac{\rho^2}{4} \left[ \Delta^{-2} \left(
\Delta^2 + 9 \ell^{-2} \right)^2 \left( dz' + \alpha \right)^2 +
\left( \Delta^2 + 9 \ell^{-2} \right) 2 \lambda^2 dw d\bar{w} \right],
\ee
where we have defined new coordinates $z'$ and $\rho$ by
\be
 z' = z - \frac{3 \log (r/\ell) }{\ell \left( \Delta^2 + 9 \ell^{-2} \right)},
\qquad \rho = 2 \left( \frac{ \Delta r}{\Delta^2 + 9 \ell^{-2}} \right)^{1/2}.
\ee
The K\"ahler form is
\be
 X^1 = \frac{1}{4} \Delta^{-1} \left( \Delta^2 + 9 \ell^{-2} \right) d
\left[ \rho^2 \left( dz' + \alpha \right) \right].
\ee
Explicitly, the base space metric for the solution (\ref{eqn:sol1})
($\Delta = \sqrt{3}/\ell$) is
\be
 ds_4^2 = d\rho^2 +  \frac{12 \rho^2}{\ell^2} \left[ dz' +
\frac{\sqrt{3}}{2\ell} \left( ydx - xdy \right) \right]^2 +
\frac{3 \rho^2}{\ell^2} \left( dx^2 + dy ^2 \right),
\ee
for the solution (\ref{eqn:sol2}) (with $0<\Delta < \sqrt{3}/\ell$) it is
\be
\label{eqn:base2}
 ds_4^2 = d\rho^2 + \frac{\rho^2}{4} \left[ \Delta^{-2} \left(9 \ell^{-2}
+  \Delta^2  \right)^2 \left( dz' + \frac{\Delta dy}{\left( 3
\ell^{-2} - \Delta^2 \right) x} \right)^2 + \left(\frac{ 9 \ell^{-2} +
\Delta^2}{3 \ell^{-2} - \Delta^2} \right) \frac{dx ^2 + dy^2}{x^2}
\right],
\ee
and for the solution (\ref{eqn:sol3}) ($\Delta > \sqrt{3}/\ell$) it is
\be
\label{eqn:base3}
 ds_4^2 = d\rho^2 + \frac{\rho^2}{4} \left[ \left( \frac{\Delta^2 + 9
 \ell^{-2}}{\Delta^2 - 3 \ell^{-2} } \right)^2 \left( d\phi + \cos \theta
 d\psi \right)^2 + \left( \frac{\Delta^2 + 9
 \ell^{-2}}{\Delta^2 - 3 \ell^{-2} } \right) \left( d\theta^2 + \sin^2 \theta
 d\psi^2 \right) \right],
\ee
where $\phi$ is defined by
\be
 z' = \frac{\Delta}{\Delta^2 - 3\ell^{-2}} \phi.
\ee
In the final case it is also useful to record
\be
 X^1 = \frac{1}{4} \left( \frac{\Delta^2 + 9 \ell^{-2}}{\Delta^2 -3
\ell^{-2}} \right) d\left[ \rho^2 \left( d\phi + \cos \theta d\psi
\right) \right].
\ee
In each case, the base space has a metric of cohomogeneity one, with a
curvature singularity at $\rho=0$. This is a good illustration of the
point that a nakedly singular base space can correspond to a
non-singular five-dimensional spacetime. Note that the metric
(\ref{eqn:base2}) has a regular limit as $\Delta \rightarrow 0$ (define
$z' = \Delta z''$) but this does {\it not} correspond to setting
$\Delta = 0$ in the solution (\ref{eqn:sol2}) (since that solution would
be null). The metric (\ref{eqn:base3}) reduces to flat space in the
limit $\Delta \rightarrow \infty$.

\sect{Supersymmetric black holes}

\label{sec:blackhole}

\subsection{Derivation of the solution}

We have determined four possibilities for the near-horizon geometry
of a supersymmetric solution with a horizon. It is natural to ask
whether any of these can arise as the near-horizon geometry of a
supersymmetric black hole. The solutions (\ref{eqn:sol1}),
(\ref{eqn:sol2}) and (\ref{eqn:sol4})
seem unlikely to arise from a solution that is asymptotically AdS (in
the global sense) since their horizons are most naturally interpreted
as spatially non-compact. However, the solution (\ref{eqn:sol3})
naturally has a horizon of $S^3$ topology so it is natural to ask
whether there is a corresponding black hole solution. In this section
we shall obtain such solutions explicitly.

We shall start by deducing some properties of the base space
associated with such a solution. Since the solution should be
asymptotically $AdS_5$, we shall demand that the base space should
asymptotically approach the base space of the $AdS_5$ solution. In
\cite{gauntlett:03} it was shown that the $AdS_5$ solution can be
obtained by talking the base space to be the Bergmann
manifold. However, $AdS_5$ is maximally supersymmetric and it might
therefore be possible to write the metric in the form (\ref{eqn:metric})
in more than one way (for example this happens for maximally
supersymmetric solutions of the ungauged theory
\cite{gauntlett:02}). In the Appendix, we show that this is not the
case: the only way of writing $AdS_5$ as a timelike solution is to use
the Bergmann manifold as the base space. The metric of the Bergmann
manifold is
\be
\label{eqn:bergmann}
 ds_4^2 = d\rho^2 + \frac{\ell^2}{4} \sinh^2 (\rho/\ell) \left[
(\sigma_L^1)^2+(\sigma_L^2)^2 \right] + \frac{\ell^2}{4} \sinh^2 (\rho/\ell)
\cosh^2 (\rho/\ell) (\sigma_L^3)^2,
\ee
where $\sigma^{i}_L$ are right-invariant $1$-forms on $SU(2)$. These can
be expressed in terms of Euler angles $(\theta,\psi,\phi)$ as
\ba
\label{eqn:sigmadef}
 \sigma_L^{1} &=& \sin \phi d\theta - \cos \phi \sin \theta d\psi \nn
 \sigma_L^{2} &=& \cos \phi d\theta + \sin \phi \sin \theta d\psi \\
 \sigma_L^{3} &=& d\phi + \cos \theta d\psi. \nonumber
\ea
where $SU(2)$ is parametrized by taking $0 \leq \theta \leq \pi$, $0
\leq \phi \leq 4 \pi$ and $0 \leq \psi \leq 2 \pi$.
The right-invariant 1-forms obey
\be
 d\sigma_L^{i} = -\frac{1}{2} \epsilon_{ijk} \sigma_L^{j} \wedge
\sigma_L^{k}.
\ee
The K\"ahler form is
\be
 X^1 = \frac{\ell^2}{4} d \left[ \sinh (\rho/\ell) \sigma_L^{3} \right].
\ee
Next we demand that near the horizon, the base space should approach
the base space of the solution (\ref{eqn:sol3}), i.e., it should agree
with equation (\ref{eqn:base3}) at small $\rho$. Now (\ref{eqn:base3}) can
be rewritten as
\be
\label{eqn:base3a}
 ds_4^2 = d\rho^2 + \frac{\rho^2}{4} \left[
\left( \frac{\Delta^2 + 9 \ell^{-2}}{\Delta^2 - 3 \ell^{-2} } \right) \left(
(\sigma_L^{1})^2 + (\sigma_L^{2})^2 \right) + \left( \frac{\Delta^2 + 9
 \ell^{-2}}{\Delta^2 - 3 \ell^{-2} } \right)^2 \left( \sigma_L^{3} \right)^2
\right],
\ee
with K\"ahler form
\be
 X^1 = \frac{1}{4} \left( \frac{\Delta^2 + 9 \ell^{-2}}{\Delta^2 -3
\ell^{-2}} \right) d\left[ \rho^2 \sigma_L^{3} \right].
\ee
In summary, we need a base space that has an asymptotic region in
which it resembles the large $\rho$ behaviour of (\ref{eqn:bergmann})
and another region in which it resembles the (singular)
small $\rho$ behaviour of
(\ref{eqn:base3a}). This suggests that try the following cohomogeneity
one ansatz for the base space of a black hole solution
\be
 ds_4^2 = d\rho^2 + a(\rho)^2 \left(
(\sigma_L^{1})^2 + (\sigma_L^{2})^2 \right) + b(\rho)^2 (\sigma_L^{3})^2,
\ee
with K\"ahler form
\be
 X^1 = d \left[ c(\rho) \sigma_L^{3} \right].
\ee
The surfaces of constant $\rho$ are homogeneous, with a transitively
acting $U(1)_L \times SU(2)_R$ isometry group, The $U(1)_L$
generated by $\partial/\partial \phi$ is manifestly a symmetry and
the $SU(2)_R$ is a symmetry because $\sigma_L^{i}$ is invariant under
the right action of $SU(2)$.

We shall assume $a,b>0$ and introduce
an orthonormal basis
\be
 e^0 =  d\rho, \, e^1 = a \sigma^{1}, \, e^2 = a \sigma^{2}, \, e^3
= b \sigma^{3}
\ee
with volume form $e^0 \wedge e^1 \wedge e^2 \wedge e^3$. The
requirement that $X^1$ be an anti-self dual complex structure reduces to
\be
\label{eqn:bc}
 c = -\epsilon a^2, \qquad b = 2 a a',
\ee
where $\epsilon = \pm 1$. We then have
\be
 X^1 = -\epsilon \left( e^0 \wedge e^3 - e^1 \wedge e^2 \right).
\ee
The base space is now determined up to one arbitrary function, namely
$a(\rho)$. The boundary conditions deduced above imply that we need
$a$ to be proportional to $\rho$ as $\rho \rightarrow 0$ and
proportional to $\exp (\rho/\ell)$ as $\rho \rightarrow \infty$.

We can write down the solution for $f$ using equation (\ref{eqn:fsol}):
\be
\label{eqn:fsol2}
 f^{-1} = {\ell^2 \over 12 a^2 a'} \big(4 (a')^3+7aa'a''-a'+a^2 a''' \big) \ .
\ee
Next we have to find a $1$-form $\omega$ on ${\cal B}$ obeying
equations (\ref{eqn:Geq}) and (\ref{eqn:maxwell}). Once again, we shall obtain a natural ansatz
for $\omega$ by examining $AdS_5$ and the near-horizon
geometry. The $AdS_5$ solution has \cite{gauntlett:03}
\be
\label{eqn:omegafar}
 \omega = \frac{\ell}{2} \sinh^2 (\rho/\ell) \sigma_L^{3}.
\ee
(At first it is a little surprising that a static solution such as
$AdS_5$ should have non-zero $\omega$ but it turns out that the
construction of \cite{gauntlett:03} yields $AdS_5$ in non-static
coordinates.) The near-horizon solution (\ref{eqn:sol3}) can
be written in the form (\ref{eqn:metric}) if we define
\be
 t = u + \frac{1}{\left( \Delta^2 + 9 \ell^{-2} \right) r},
\ee
so $V = \partial/\partial t = \partial/\partial u$, and
\be
\label{eqn:omeganear}
 \omega = \frac{3 \sigma_L^{3}}{\Delta \ell \left( \Delta^2 - 3 \ell^{-2}
\right) r} = \frac{12 \sigma_L^{3}}{\ell \left( \Delta^2 - 3\ell^{-2} \right)
\left( \Delta^2 + 9 \ell^{-2} \right) \rho^2}.
\ee
If a supersymmetric black hole solution exists then it should have
$\omega$ that behaves as (\ref{eqn:omegafar}) for large $\rho$ and as
(\ref{eqn:omeganear}) for small $\rho$. This suggests the Ansatz
\be
 \omega = \Psi(\rho) \sigma_L^{3},
\ee
which gives
\be
\label{eqn:gplusbh}
 G^\pm = \frac{f}{2} \left( \frac{\Psi'}{2aa'} \mp \frac{\Psi}{a^2}
\right) \left( e^0 \wedge e^3 \pm e^1 \wedge e^2 \right).
\ee
It remains to solve equations (\ref{eqn:Geq}) and (\ref{eqn:maxwell}).
Equation (\ref{eqn:Geq}) reduces to
\be
\label{eqn:psi1}
 \frac{\Psi'}{2aa'} - \frac{\Psi}{a^2} =  \frac{\epsilon \ell g}{2 f}
\ee
where
\be
 g = -\frac{a'''}{a'} - 3\frac{a''}{a} - \frac{1}{a^2} +
 4\frac{{a'}^2}{a^2}.
\ee
Using equation (\ref{eqn:psi1}) we can write (\ref{eqn:maxwell}) as
\be
\label{eqn:psi2}
 \frac{\Psi'}{2aa'} + \frac{\Psi}{a^2} = -\frac{\epsilon \ell}{2} \left(
 \nabla^2 f^{-1} + 8 \ell^{-2} f^{-2} - \frac{\ell^2 g^2}{18} \right).
\ee
Note that
\be
 \nabla^2 f^{-1} = \frac{1}{a^3 a'} \partial_\rho \left(a^3 a'
\partial_\rho f^{-1} \right).
\ee
Eliminating $\Psi'$ between equations (\ref{eqn:psi1}) and
(\ref{eqn:psi2}) gives
\be
\label{eqn:psisol}
 \Psi = -\frac{\epsilon \ell a^2}{4} \left( \nabla^2 f^{-1} + 8 \ell^{-2}
 f^{-2} -\frac{\ell^2 g^2}{18} + f^{-1} g \right).
\ee
This expression uniquely determines $\Psi$ in terms of
$a$. Substituting this back into equation (\ref{eqn:psi1}) or
(\ref{eqn:psi2}) then gives
\be
\label{eqn:ode}
 \left( \nabla^2 f^{-1} + 8 \ell^{-2}
 f^{-2} -\frac{\ell^2 g^2}{18} + f^{-1} g \right)' + \frac{4 a' g}{a f}
= 0,
\ee
which is a sixth order nonlinear ordinary differential equation in
$a(\rho)$. Everything else is determined once we have solved this
equation. We are already familiar with two solutions of this equation,
namely those corresponding to the $AdS_5$ and near-horizon
geometries. Now we are looking for a more general solution that
interpolates between these two types of behaviour. First we need to
understand the boundary conditions at $\rho=0$. From the near horizon
geometry, we know that the coordinate transformation relating $\rho$
to $r$ must be of the form $r \propto \rho^2$ for small $\rho$. Hence
smoothness of the near-horizon geometry implies that only even powers
of $\rho$ should appear in the five dimensional metric. Looking at the
expression for $f$, this appears to imply that $a$ must contain either
only odd powers or only even powers of $\rho$. Since we need $a
\propto \rho$ for small $\rho$ we must take the former
possibility. Hence we demand $a(0)=a''(0)=a^{(4)}(0)=0$. The equation
(\ref{eqn:ode}) determines the sixth derivative of $a$ in terms of lower
derivatives. Therefore demanding that $a^{(6)}(0)=0$ gives an equation
relating $a^{(5)}(0)$ to $a'''(0)$ and $a'(0)$. Hence the solutions of
interest are determined by specifying $a'''(0)$ and $a'(0)$.

In fact by rescaling the coordinates $t = \lambda \tilde{t}$, $\rho =
\tilde{\rho}/\sqrt{\lambda}$, $a(\rho)$ is replaced by
$\tilde{a}(\tilde{\rho}) = \sqrt{\lambda}
a(\tilde{\rho}/\sqrt{\lambda})$. This has the effect of rescaling
$a'''(0)$ by $\lambda^{-1}$ but leaves $a'(0)$ invariant. Hence
$a'''(0)$ can be rescaled to any convenient value; it is only the {\it
sign} of $a'''(0)$ that is important. Hence we would naively expect to
find just three $1$-parameter familes of solutions
satisfying the boundary conditions required for a smooth horizon. The
three families will correspond to $a'''(0)$ positive, negative or
zero.

We are already familiar with a $1$-parameter family of solutions with
$a'''(0)=0$: these are just the near-horizon solutions (\ref{eqn:base3a})
with $a(\rho) \propto \rho$. They are labelled by the parameter
$\Delta$.

Another solution we know is the $AdS_5$ solution, with $a = (\ell/2)
\sinh (\rho/\ell)$. This has $a'''(0)>0$. The form of this solution
suggests that we look for a more general solution of the form
$a = \alpha \ell \sinh (\beta \rho/\ell)$. Amazingly, this turns out to be a
solution to (\ref{eqn:ode}) for arbitrary $\alpha$ and $\beta$. It is
really just a 1-parameter family of solutions since the above
rescaling can be used to set $\beta$ to any convenient value. We shall
choose $\beta=1$, i.e.,
\be
\label{eqn:asol}
 a = \alpha \ell \sinh (\rho/\ell).
\ee
This is a 1-parameter family of solutions with $a'''(0)>0$.

Finally, if we take this solution and analytically continue $\alpha
\rightarrow -i \alpha$ and $\beta \rightarrow i \beta$ we obtain the
solution $a = \alpha \ell \sin (\beta \rho /\ell)$.  Again, we can rescale
to set $\beta=1$, i.e.,
\be
 a = \alpha \ell \sin (\rho/\ell).
\ee
This is a $1$-parameter solution with $a'''(0)<0$.

Of these three families of solutions, which is most likely to describe
a black hole? By construction, all have a regular horizon at
$\rho=0$. However, to describe a black hole they must also be
asymptotically $AdS_5$. The first family just describes the
near-horizon geometry and does not have the correct asymptotics.
The third family looks as if it has problems at $\rho = \ell \pi /2$
($b=0$ there).
This leaves the second family. We shall now show that this does indeed
describe a 1-parameter family of supersymmetric black holes in $AdS_5$.

First, comparing the small $\rho$ behaviour of (\ref{eqn:asol}) with the
corresponding behaviour in (\ref{eqn:base3a}) we find that our solution
has a regular near-horizon geometry with parameter $\Delta$ determined
by
\be
 \alpha = \frac{1}{2} \sqrt{ \frac{ \Delta^2 + 9 \ell^{-2}}{\Delta^2 -
3\ell^{-2}} },
\ee
hence $\alpha > 1/2$ is necessary for the solution to have a regular
horizon. $\alpha = 1/2$ does not give a horizon: it is just the
$AdS_5$ solution in global coordinates.

Equations (\ref{eqn:bc}), (\ref{eqn:fsol2}) and (\ref{eqn:psisol}) give the
full solution as
\be
 a = \alpha \ell \sinh (\rho/\ell), \qquad b = 2 \alpha^2 \ell \sinh(\rho/\ell)
\cosh (\rho/\ell), \qquad c = -\epsilon \alpha^2 \ell^2 \sinh^2 (\rho/\ell),
\ee
\be
 f^{-1} = 1 + \frac{4 \alpha^2 -1}{12 \alpha^2 \sinh^2 (\rho/\ell)},
\ee
\be
 \Psi = -2 \epsilon \alpha^2 \ell \sinh^2 (\rho/\ell) \left[ 1 + \frac{4
\alpha^2 -1}{4\alpha^2 \sinh^2 (\rho/\ell)} + \frac{(4\alpha^2 -1)^2}{96
\alpha^4 \sinh^4(\rho/\ell)} \right].
\ee
The Maxwell field strength is obtained from equation
(\ref{eqn:maxwell1}). It can be written in terms of the potential
\be
 A = \frac{\sqrt{3}}{2} f \left[ dt + \frac{\epsilon \ell ( 4\alpha^2 -1
)^2}{ 144 \alpha^2 \sinh^2 (\rho/\ell) } \sigma_L^{3} \right].
\ee

\subsection{New coordinates}

\label{sec:newcoord}

It is convenient to introduce a new radial coordinates $R$ defined by
\be
 \frac{1}{4} R^2  =  f^{-1} a^2,
\ee
that is
\be
 R = \ell \sqrt{4 \alpha^2 \sinh^2 (\rho/\ell) + \frac{4 \alpha^2 -1
}{3}}.
\ee
Let
\be
 R_0 = \ell \sqrt{ \frac{ 4\alpha^ 2- 1}{3} }.
\ee
We then have (recall that $\epsilon=\pm 1$)
\be
 f = 1 - \frac{R_0^2}{R^2}, \qquad \Psi = -\frac{\epsilon R^2}{2\ell}
\left( 1 + \frac{2 R_0^2}{R^2} + \frac{3 R_0^4}{2 R^2 (R^2- R_0^2)}
\right).
\ee
The horizon is at $R=R_0$. The full metric is
\be
\label{eqn:fullsol}
 ds^2 =  - f^2 dt^2 - 2 f^2 \Psi dt \sigma_L^{3} + U(R)^{-1} dR^2
 + \frac{R^2}{4} \left[ (\sigma_L^{1})^2 + (\sigma_L^{2})^2 +
\Lambda(R) ( \sigma_L^{3})^2 \right],
\ee
where $\sigma_L^{i}$ was defined in equation (\ref{eqn:sigmadef}), and
\be
 U(R) = \left( 1 - \frac{R_0^2}{R^2} \right)^2 \left( 1 + \frac{2
R_0^2}{\ell^2} + \frac{R^2}{\ell^2} \right), \qquad \Lambda(R) = 1 +
\frac{R_0^6}{\ell^2 R^4} - \frac{R_0^8}{4 \ell^2 R^6}.
\ee
The Maxwell potential is
\be
 A = \frac{\sqrt{3}}{2} \left[ \left( 1 - \frac{R_0^2}{R^2} \right) dt
 + \frac{\epsilon R_0^4}{4 \ell R^2}
\sigma_L^{3} \right].
\ee
One final coordinate transformation is required to demonstrate that
the solution is asymptotically $AdS_5$. Let
\be
 \phi' = \phi + \frac{2\epsilon t}{\ell}
\ee
and
\be
 \Omega(R) = \frac{2\epsilon}{\ell \Lambda(R)} \left[ \left(\frac{3}{2} +
\frac{R_0^2}{\ell^2} \right) \frac{R_0^4}{R^4} - \left( \frac{1}{2} +
\frac{R_0^2}{4\ell^2} \right) \frac{R_0^6}{R^6} \right].
\ee
The metric becomes
\be
\label{eqn:metricasympads}
 ds^2 = -U(R) \Lambda(R)^{-1} dt^2 + U(R)^{-1} dR^2 + \frac{R^2}{4} \left(
(\sigma_L^{1'})^2 + (\sigma_L^{2'})^2 + \Lambda(R) \left( \sigma_L^{3'} -
\Omega(R) dt \right)^2 \right),
\ee
where $\sigma_L^{i'}$ is defined in the same way as $\sigma_L^{i}$
(equation (\ref{eqn:sigmadef})) but with $\phi$ replaced by $\phi'$.
The electromagnetic potential is
\be
 A = \frac{\sqrt{3}}{2} \left[ \left( 1 - \frac{R_0^2}{R^2} -
\frac{R_0^4}{2 \ell^2 R^2} \right) dt + \frac{\epsilon R_0^4}{4 \ell R^2}
\sigma_L^{3'} \right].
\ee
Now $\Lambda \rightarrow 1$, $\Omega \rightarrow 0$ and $U \sim
R^2/\ell^2$ as $R \rightarrow \infty$ so
the solution is manifestly asymptotic to $AdS_5$.
By construction it has a regular horizon at $R=R_0$. It is
well-behaved for $R>R_0$. Hence it describes a $1$-parameter family of
supersymmetric, asymptotically $AdS_5$ black holes. The isometry group
is $R \times U(1)_L \times SU(2)_R$, the same as that of the
non-static supersymmetric black holes of the ungauged theory
\cite{gauntlett:99}.

\subsection{Properties of the solution}

We have seen that our solution is asymptotically $AdS_5$. It therefore
has the Einstein universe $R \times S^3$ as its conformal boundary.
In the coordinates induced from our bulk solution
(\ref{eqn:metricasympads}), the boundary metric can be written
\be
\label{eqn:bdymetric}
 ds^2 = -dt^2 + \frac{\ell^2}{4} \left( (\sigma_L^{1'})^2 + (\sigma_L^{2'})^2 +
(\sigma_L^{3'})^2 \right).
\ee
The $S^3$ has radius $\ell$. Boundary time translations are generated by
$\partial/\partial t$ so it is natural to use $\partial/\partial t$ as
the generator of bulk time translations too. However, for rotating,
asymptotically AdS, black holes, there is a subtlety
\cite{hawking:99}: there exists another timelike Killing vector field
in the bulk. In our case, this is the Killing vector
field $V$ associated with supersymmetry. In the coordinates of
(\ref{eqn:metricasympads}), this is given by
\be
\label{eqn:angvel}
 V= \frac{\partial}{\partial t} + \frac{2\epsilon}{\ell}
 \frac{\partial}{\partial \phi'}.
\ee
Furthermore, $V$ is tangent to the event horizon of the black
hole so using $V$ as the generator of time translations
corresponds to working in a co-rotating frame.
The advantage of using $V$ as the generator of time
translations is that $V$ is timelike everywhere outside the black
hole, whereas $\partial/\partial t$ becomes spacelike near the event
horizon. Hence if $\partial/\partial t$ generates time translations
then there is an ergoregion, but if the co-rotating
Killing vector field $V$ generates time translations then there is no
ergoregion. This is a general feature of rotating, asymptotically AdS,
black holes \cite{hawking:00}.

Note that $V$ is null on the conformal boundary with metric
(\ref{eqn:bdymetric}). In this sense, the boundary rotates at the
speed of light in the co-rotating frame. The same is true of
supersymmetric rotating black holes in $D=3$ \cite{hawking:99}
and $D=4$ \cite{caldarelli:99}. Conversely, if we take
$\partial/\partial t$ as the generator of time translations then
equation \ref{eqn:angvel} shows that the angular velocity of the black
hole in the $\phi'$ direction is
\be
 \Omega_H = \frac{2\epsilon}{\ell},
\ee
which implies that the black hole rotates at the speed of light with
respect to the stationary frame at infinity.

We are ultimately interested in the AdS/CFT description of these black
holes. From a CFT perspective, it seems most natural to use
$\partial/\partial t$ (in the coordinates (\ref{eqn:metricasympads})) as
the generator of time translations so we shall do that henceforth.

Ashtekar and Das (AD) have shown how to define conserved quantities in
asymptotically anti-de Sitter spacetimes of arbitrary dimension $D \ge
4$ \cite{ashtekar:00}. There is a conserved quantity associated with each
symmetry of the conformal boundary.
The AD mass is the conserved quantity associated with
$\partial/\partial t$ and takes the value
\be
 M = \frac{3\pi R_0^2}{4 G} \left( 1 + \frac{3 R_0^2}{2\ell^2} + \frac{2
R_0^4}{3 \ell^4} \right).
\ee
We define the angular momentum to be {\it minus}\footnote{
The minus sign is required to give the correct sign for the
angular momentum of known black hole solutions, such as those of
\cite{hawking:99}. It is the same relative minus sign as occurs in the
definition of conserved quantities for geodesics: if $U$ is the
tangent vector to a geodesic then the energy is $E = - U \cdot
\partial/\partial t$ and the angular momentum is $L = + U \cdot
\partial/\partial \phi$.}
the conserved quantity associated with $\partial/\partial \phi'$,
which gives
\be
 J = \frac{3 \epsilon \pi R_0^4}{8 G \ell} \left( 1 + \frac{2 R_0^2}{3
\ell^2} \right),
\ee
and the angular momentum associated with $\partial/\partial \psi$
vanishes. Hence our solutions carry equal angular momenta in two
orthogonal $2$-planes, just like the supersymmetric black holes of the
ungauged theory \cite{breckenridge:97,gauntlett:99}. Note that the
choice of the sign $\epsilon$ fixes the sign of $J$.

An alternative definition of conserved charges is provided by the
``holographic stress tensor" approach \cite{balasubramanian:99,
kraus:99, deharo:01, skenderis:01}. In this method, the expectation
value of the stress tensor of the dual CFT on $R \times S^3$, in a
conformal frame with metric (\ref{eqn:bdymetric}), is calculated using the
formula
\be
 \langle T_{\mu\nu} \rangle = \lim_{R \rightarrow \infty} \frac{R^2}{8
 \pi G \ell^2} \left[ - \left( K_{\mu\nu} - K h_{\mu\nu} \right) -
 \frac{3}{\ell} h_{\mu\nu} + \frac{\ell}{2} \left( \bar{R}_{\mu\nu} -
 \frac{1}{2} \bar{R} h_{\mu\nu} \right) \right].
\ee
The right hand side is defined in terms of tensors associated with a
surface $R={\rm constant}$ in the coordinates of
(\ref{eqn:metricasympads}). This surface has induced metric
$h_{\mu\nu}$, extrinsic curvature $K_{\mu\nu}$ with trace $K$, Ricci
tensor $\bar{R}_{\mu\nu}$ and Ricci scalar $\bar{R}$. An overall factor of
$R^2/\ell^2$ arises from the conformal transformation required to
map $h_{\mu\nu}$ to the metric (\ref{eqn:bdymetric}) as $R \rightarrow
\infty$. We obtain
\bea
 8 \pi G \langle T_{\mu\nu} \rangle dx^{\mu} dx^{\nu} &=&
 \frac{1}{\ell} \left( \frac{3}{8} + \frac{3 R_0^2}{\ell^2} + \frac{9
 R_0^4}{2\ell^4} + \frac{ 2R_0^6}{\ell^6} \right) dt^2 - 2\epsilon
 \left( \frac{3 R_0^4}{2 \ell^4} + \frac{R_0^6}{\ell^6} \right) dt
 \sigma_L^{3'} \nn
 &+& \frac{\ell}{32} \left( 1 + \frac{8R_0^2}{\ell^2} + \frac{12
 R_0^4}{\ell^4} \right) \left[ (\sigma_L^{1'})^2 + (\sigma_L^{2'})^2
 \right] \\
 &+& \frac{\ell}{32} \left( 1 + \frac{8R_0^2}{\ell^2} + \frac{12
 R_0^4}{\ell^4} + \frac{16 R_0^6}{\ell^6} \right) (\sigma_L^{3'})^2.\nonumber
\eea
The total energy $E$ is obtained by integrating $\langle T_{tt} \rangle$ over $S^3$,
giving
\be
 E = M + \frac{3 \pi \ell^2}{32 G}.
\ee
The final term is just the Casimir energy of the CFT on $R \times
S^3$, which is interpreted as the energy of global $AdS_5$ in this
approach \cite{balasubramanian:99}. The total angular momentum is
obtained by integrating $-\langle T_{t\phi'} \rangle$ over $S^3$,
which gives the same result $J$ as the AD method used above.

Define the electric charge by
\be
 Q = \frac{1}{4\pi G} \int_{S^3_\infty} \star F,
\ee
where the integral is over the three-sphere at infinity on a surface
of constant $t$. The orientation of spacetime can be deduced from the
orientation of the base space to be $dt \wedge dR \wedge
\sigma_L^1 \wedge \sigma_L^2 \wedge \sigma_L^3$. For our solution we
find\footnote{
One might wonder why this is always non-negative. This is related to
our choice $f>0$. Solutions with $f<0$ are related by a change of
orientation \cite{gauntlett:02}, and will have the opposite sign for $Q$.}
\be
 Q =  \frac{ \sqrt{3}  \pi R_0^2}{2G} \left( 1 +
\frac{R_0^2}{2 \ell^2} \right).
\ee
We then have
\be
 M - \frac{2}{\ell} |J| = \frac{\sqrt{3}}{2} |Q|.
\ee
Note that the left hand side is the AD conserved quantity
associated with $V$.
The BPS bound for this theory is \cite{london:95}\footnote{
The final result presented in \cite{london:95} assumed $J_1 = J_2 = 0$
but it is straightforward to use the intermediate results of \cite{london:95} to take
account of non-zero $J_i$.}
\be
 M - \frac{|J_1|+|J_2|}{\ell} \ge \frac{\sqrt{3}}{2} |Q|
\ee
where $J_{1,2}$ are the angular momenta associated with the orthogonal
$2$-planes at infinity. Our black hole saturates this inequality with
$J_1 = J_2 = J$. Although we derived our solution within the framework
of \cite{gauntlett:03}, the quickest way of checking it is to
verify that it satifies the equations of motion. It must then be
supersymmetric because it saturates the BPS bound.

Our solution has a regular horizon by construction. To find out what
lies behind the horizon, we transform to Gaussian null coordinates as
follows:
\be
 dt = du - \Lambda U^{-1} dr, \qquad d\phi = d\phi'' - \frac{4 f^2
\Psi}{R^2 U} dr,  \qquad dR = \Lambda^{1/2} dr,
\ee
with $R=R_0$ at $r=0$. In these coordinates, the metric is
\be
 ds^2 = - f^2 du^2 + 2 du dr - 2 f^2 \Psi du \sigma_L^{3''} + \frac{R^2}{4} \left[
(\sigma_L^{1''})^2 + (\sigma_L^{2''})^2 + \Lambda ( \sigma_L^{3''})^2 \right],
\ee
where we have used $UR^2 = f^2(4 f^2 \Psi^2  + R^2 \Lambda)$, and
$\sigma_L^{i''}$ is defined in the same way as $\sigma_L^{i}$
(equation (\ref{eqn:sigmadef})) but with $\phi$ replaced by
$\phi''$. Note that $\Lambda$ approaches a non-zero constant at $R=R_0$
so $R-R_0$ is proportional to $r$ there. It is then easy to see that
$f^2 \propto r^2$ and $f^2 \Psi \propto r$ as $r \rightarrow 0$, so
the above metric is of the form (\ref{eqn:metric2}), with a smooth event
horizon at $r=0$. A gauge transformation $A \rightarrow A + d\lambda$
with $\lambda=\lambda(r)$ makes $A$ regular at the horizon with $A$ a
constant multiple of $\sigma_L^{3''}$ on the horizon.
It is straightforward to read off the spatial geometry of the horizon:
\be
 ds_3^2 = \frac{R_0^2}{4} \left[ (\sigma_L^{1''})^2 + (\sigma_L^{2''})^2 +
 \left( 1 + \frac{3R_0^2}{4 \ell^2} \right) (\sigma_L^{3''})^2 \right],
\ee
a squashed $S^3$. This is no surprise because the black hole
was constructed to have near-horizon geometry (\ref{eqn:sol3}). The area
of the horizon is
\be
 {\cal A} = 2\pi^2 R_0^3 \sqrt{ 1 + \frac{3 R_0^2}{4 \ell^2} }.
\ee

In the region $r<0$ behind the horizon, we can invert the above
coordinate transformation to return to the original coordinates
$(t,R,\theta,\psi,\phi)$. The metric is just (\ref{eqn:fullsol}) with
$R<R_0$. This metric is smooth for $R>0$ but has a curvature
singularity at $R=0$. Note that there is a value $R=R_*$, $0< R_* <
R_0$ with $\Lambda(R_*)=0$. Hence $\partial/\partial \phi$ is null at
$R=R_*$. However, $\phi$ is periodic hence there are closed null
curves. Similarly, for $R<R_*$ there are closed timelike curves. Hence
the interior of the black hole contains a curvature singularity
surrounded by a region of closed timelike curves.

In the Gaussian null coordinates, a curve of constant $u$, $\theta$,
$\psi$ and $\phi''$ with $r=-\lambda$ is a future-directed null
geodesic with affine parameter $\lambda$. Let $r_*<0$ denote the
value of $r$ at which $R=R_*$. It is easy to see that
\be
 R-R_* \approx \frac{1}{4} \Lambda'(R_*) (r-r_*)^2.
\ee
Hence, as $r$ decreases through $r_*$, $R$ decreases to $R_*$ but then
starts to increase again. As $r$ decreases further, $R$ will increase
through $R_0$ so this null geodesic will eventually emerge into
another asymptotically $AdS$ region.

These considerations suggest that
the global causual structure of our solution is very similar to that
of the supersymmetric rotating black holes of the ungauged theory,
which was explored in detail  in \cite{gibbons:99}. The only qualitative
difference is that our solution is asymptotically $AdS$ rather than
asymptotically flat.

Finally, we note that if we take the limit $\ell \rightarrow \infty$
with $R_0$ held fixed then our solution reduces to a {\it static}
supersymmetric black hole solution of the ungauged supergravity
theory. The four dimensional solutions of \cite{kostalecky:96} behave
similarly.

\subsection{A limiting case}

We shall now examine the geometry of our solution when the black hole
is very large, i.e., $R_0 \rightarrow \infty$. To this end, it is
convenient to write $\sigma_L^{3}$ as
\be
 \sigma_L^{3} = (d\phi + d\psi) - 2 \sin^2 \frac{\theta}{2} \,
d\psi,
\ee
and define new coordinates $S$, $\tilde{\theta}$, $z$
\be
 S = \frac{R}{R_0}, \qquad \tilde{\theta} = \frac{1}{2} R_0 \theta,
\qquad z = \frac{\sqrt{3} R_0^2}{4\ell} ( \phi + \psi ).
\ee
We then take the limit $R_0 \rightarrow \infty$ holding $t$, $S$,
$\tilde{\theta}$, $z$ and $\psi$ fixed. If we define
\be
 x = \tilde{\theta} \cos \psi, \qquad y = \tilde{\theta} \sin \psi,
\ee
then the limiting form of the solution is
\ba
\label{eqn:limit}
 ds^2 &=& - \left( 1 - S^{-2} \right)^2 \left[ dt^2 - \frac{ 4
\epsilon}{\sqrt{3}}  \left(S^2 + 2 +
\frac{3}{2(S^2-1)} \right) dt  \left( dz + \frac{\sqrt{3}}{2\ell} (y dx-xdy)
\right) \right] \\ &+& \frac{\ell^2 dS^2}{ \left( 1 - S^{-2} \right)^2 \left( 2 + S^2
\right)}
 + S^2 \left( dx^2 + dy^2 \right) + \frac{1}{3} \left( 4 S^{-2} - S^{-4}
\right) \left( dz + \frac{\sqrt{3}}{2\ell} (y dx-x dy) \right)^2, \nn
 A &=& \frac{\sqrt{3}}{2} \left[ \left( 1 - S^{-2} \right) dt +
\frac{\epsilon}{\sqrt{3} S^2} \left( dz + \frac{\sqrt{3}}{2\ell} (y dx
-x  dy)
 \right) \right].\nonumber
\ea
By transforming to Gaussian null coordinates in the manner described
above, it is easy to see that this solution has a regular horizon at
$S=1$, with near-horizon geometry (\ref{eqn:sol1}). Thus taking this
limit has the effect of performing a group contraction of $H$ from
$S^3$ to $Nil$.

This solution is not asymptotically $AdS$: as $S \rightarrow \infty$,
it tends to the following vacuum solution
\ba
\label{eqn:limitasymp}
 ds^2 &=& - \left[dt - \frac{2 \epsilon S^2}{\sqrt{3}} \left( dz +
\frac{\sqrt{3}}{2\ell} (y dx-x dy) \right) \right]^2
\nn &+& \frac{\ell^2 dS^2}{S^2} + S^2 \left( dx^2 + dy^2
\right) + \frac{4 S^4}{3} \left( dz + \frac{\sqrt{3}}{2\ell} (y dx - x dy)
\right)^2, \nn
F &=& 0.
\ea
It can be verified that this describes a supersymmetric solution in
its own right. Note that, although this solution belongs to the
timelike family, $\partial/\partial z$ is a null Killing
vector field. Null geodesics tangent to $\partial/\partial z$ are free
of expansion, rotation and shear so this solution is a plane-fronted
wave.

\sect{Discussion}

\label{sec:discussion}

It is natural to ask whether our 1-parameter family of solutions
is a special example in some larger family of supersymmetric, asymptotically
$AdS_5$ black hole solutions. There are two reasons why this seems
unlikely. First, if there was a more general family of solutions then
presumably there would be a more general family of near-horizon
solutions than \ref{eqn:sol3}. This more general family would have to
have non-constant $\Delta$ on the horizon. We are not aware of any
black hole solution in any dimension for which $\Delta$ is
non-constant. Secondly, the results of \cite{kostalecky:96} show that
in $D=4$ there is at most one supersymmetric black hole for given
charge(s). Our $D=5$ solution has the same property, but this would
not be true if it were part of some larger family.

Of course, we would expect more general solutions to exist in more
complicated theories. Of particular interest is the maximal ${\cal N}=4$
$D=5$ gauged supergravity (i.e., 32 supercharges) with $SO(6)$ gauge
group arising from Kaluza-Klein reduction of type IIB supergravity
on $S^5$. This can be truncated to a ${\cal N}=1$ theory with $U(1)^3$ gauge
symmetry. When embedded in this theory, our solutions have equal $U(1)$
charges but it should be straightforward to generalize them to obtain
solutions with three independent charges.
The form of our solution suggests a natural Ansatz
for the metric and gauge fields of such solutions, and one would
assume the scalar fields to be functions of $R$ alone.
It would also
be interesting to find non-extremal generalizations of our
solution. These should form a 4-parameter family of solutions of
minimal $D=5$ gauged supergravity, parametrized by the mass, charge
and two independent angular momenta. They should reduce to the
solutions of \cite{hawking:99} in the limit of vanishing charge.

Another interesting question concerns black hole uniqueness.
General stationary asymptotically flat $D=5$ black holes
do not obey a uniqueness theorem \cite{emparan:02}. However,
supersymmetric black holes do \cite{reall:03}, at least in minimal
supergravity. It is not known whether similar behaviour occurs for
asymptotically AdS black holes. It seems unlikely that the proof of
\cite{reall:03} could be extended to gauged supergravity. The first
problem is in classifying all near-horizon geometries. We were
unable to do this without additional assumptions. However, it might be
possible to perform a complete analysis. The second problem is more
serious, namely that the proof relies on a global property of
hyper-K\"ahler manifolds that has no K\"ahler analogue.

The near-horizon solutions that we found preserve at least $1/4$
supersymmetry. In the ungauged theory, the
near-horizon geometry of a supersymmetric black hole is actually
maximally supersymmetric \cite{gauntlett:99}. This cannot be true in
the gauged theory because the only maximally supersymmetric solution is
$AdS_5$ \cite{gauntlett:03}. However, the near-horizon solutions might
be $1/2$ or $3/4$ supersymmetric, as seems to be the case with the $AdS_3
\times H^2$ solution. It might be possible to extend the work of
\cite{gauntlett:03} to classify all $1/2$ and $3/4$ supersymmetric solutions of
this theory.

The main objective of our work was to stimulate further
investigation of black holes using the AdS/CFT correspondence. Supersymmetric,
asymptotically AdS, black holes have been studied as limits of
non-supersymmetric solutions in $D=3$ \cite{hawking:99} and $D=4$
\cite{caldarelli:99}. The matter of the dual CFT rotates around the
Einstein universe with a speed that approaches the speed
of light as the supersymmetric limit is taken. It seems very likely
that the same will be true of the CFT states corresponding to our
solution. (As we have seen, the boundary rotates at the speed of light
with respect to the co-rotating frame in the bulk.)
We hope that by counting such states
(with appropriate R-charge), it will be possible to reproduce the
entropy of our solutions.

It would also be interesting
to obtain a microscopic understanding of these black holes from a bulk
perspective. Our solution can be embedded into type IIB supergravity
using \cite{chamblin:99}. The only non-trivial fields are the metric
and 5-form, which suggests that the bulk description should only involve
D3-branes. The charge of our solutions is Kaluza-Klein charge arising
from momentum on the internal $S^5$. BPS solutions describing
probe D3-branes moving on the internal $S^5$ of $AdS_5 \times S^5$ are
known: giant gravitons \cite{mcgreevy:00}. Going beyond
the probe approximation, one can construct supersymmetric solutions
of minimal $D=5$ gauged supergravity by considering large numbers of
giant gravitons suitably distributed on $S^5$ \cite{myers:02}. These
solutions correspond to static point sources in $AdS_5$, and are
therefore nakedly singular. However, as we have shown, the
inclusion of angular momentum can lead to a regular horizon.
Giant gravitons that carry angular momentum both on $S^5$ and in
$AdS_5$ were investigated in \cite{arapoglu:03}. Perhaps our
solutions can be interpreted as a distribution of such objects.

\acknowledgments{
This research was supported in part by the National Science Foundation
under Grant No. PHY99-07949. J. B. G. was supported by EPSRC. We thank
Jerome Gauntlett and Kostas Skenderis for useful correspondence.}

\appendix
\makeatletter
\renewcommand{\theequation}{A.\arabic{equation}}
\@addtoreset{equation}{section} \makeatother
\section{Maximally Symmetric Solutions}

In \cite{gauntlett:03} it was shown that by taking the K\"ahler base space
with a Bergmann metric, one can obtain the maximally symmetric $AdS_5$ geometry in
five dimensions. Here we shall prove a (partial) converse, namely that
the only maximally symmetric timelike solution of five dimensional
minimal gauged supergravity is $AdS_5$ (with $AdS$ radius $\ell$), and
that the base space is locally isometric to the Bergmann manifold.

To show this, suppose that
\be
\label{eqn:maxsym}
{}^5 R_{\alpha \beta \rho \sigma} = {\lambda \over 4} (g_{\beta
\rho}g_{\alpha \sigma}-g_{\alpha \rho} g_{\beta \sigma})
\ee
for some constant $\lambda$. Hence, in particular, ${}^5 R_{\alpha
\beta} = - \lambda g_{\alpha \beta}$. So, the Einstein equations imply that
\be
F^2 = {60 \over \ell^2} -15 \lambda
\ee
and hence
\be
\label{eqn:apa}
3 g_{\alpha \beta} ({4 \over \ell^2} - \lambda)= F_{\alpha \gamma}F_{\beta}{}^\gamma
\ee
If $\lambda \neq {4 \over \ell^2}$ then ({\ref{eqn:apa}}) implies that
the determinant of the metric vanishes. To prevent this, we must have
$\lambda ={4 \over \ell^2}$, so the five-dimensional geometry is
$AdS_5$ with $AdS$ radius $\ell$. From $F_{\alpha
\gamma}F_{\beta}{}^\gamma=0$ we obtain $F=0$.

Using the identity for $F$ given in ({\ref{eqn:maxwell}}) we see that $f$ must be constant.
Without loss of generality we can set $f=1$. In addition,
\bea
G^+ &=& 0
\nn
d \omega &=& {2 \over \ell} X^1
\eea
Note that the only non-vanishing components of the spin connection are fixed by
\bea
\omega_{0ij} &=& \omega_{i0j} ={1 \over \ell} X^1{}_{ij}
\nn
\omega_{ijk} &=& {\hat{\omega}}_{ijk}
\eea
where here ${\hat{\omega}}_{ijk}$ denotes the spin connection of the K\"ahler base, and
$i,j,k$ are frame indices with respect to a vielbein on the base. Using these expressions it
is straightforward to compute the following components of the Riemann tensor
\be
{}^5R_{ijpq} = {}^4 R_{ijpq}+{2 \over \ell^2} \big( X^1{}_{ij} X^1{}_{pq} +{1 \over 2}(
 X^1{}_{pi} X^1{}_{qj} - X^1{}_{qi} X^1{}_{pj}) \big)
\ee
However, from ({\ref{eqn:maxsym}}) we must have
\be
{}^5 R_{ijpq} = {1 \over \ell^2} (\delta_{jp} \delta_{iq} - \delta_{ip} \delta_{jq})
\ee
and so solving for ${}^4 R_{ijpq}$ we obtain
\be
{}^4 R_{ijpq} = {2 \over \ell^2} \big(-X^1{}_{ij} X^1{}_{pq} +{1 \over 2}(
 X^1{}_{qi} X^1{}_{pj}- X^1{}_{pi} X^1{}_{qj}) -{1 \over 2}(\delta_{ip} \delta_{jq}
 -\delta_{jp} \delta_{iq}) \big)
\ee
This expression implies that the K\"ahler base space has constant
holomorphic sectional curvature, which must be the same as that of the
Bergmann manifold (since we know that is a solution to these equations).
However, any two K\"ahler manifolds with the same constant holomorphic
sectional curvature are locally holomorphically isometric
\cite{kobayashi:96}. Hence the base space must be locally
holomorphically isometric to the Bergmann manifold.

\end{document}